\documentclass{article}

\def\xxinput#1{\input#1}

\xxinput{vsolj02.sty}

\usepackage[dvipdfmx]{graphicx}
\usepackage[comma,colon]{natbib}
\usepackage[hidelinks,breaklinks]{hyperref}

\usepackage[OT2,T1]{fontenc}

\def\cite{\citealt}
\setcitestyle{aysep={}}

\newcounter{author}
\def\altaffilmark#1{$^{#1}$}
\def\altaffiltext#1{$^{#1}$\,}

\def\authorcount#1#2{{\refstepcounter{author}\label{#1}
                     \altaffiltext{\ref{#1}}{#2}}}

\begin{document}

\begin{center}

\title{Genuine standstill in the AM CVn star CR Boo}

\author{
        Taichi~Kato\altaffilmark{\ref{affil:Kyoto}}
        Yutaka~Maeda,\altaffilmark{\ref{affil:Mdy}}
        Masayuki~Moriyama\altaffilmark{\ref{affil:Myy}}}
\email{tkato@kusastro.kyoto-u.ac.jp}

\authorcount{affil:Kyoto}{
     Department of Astronomy, Kyoto University, Sakyo-ku,
     Kyoto 606-8502, Japan}

\authorcount{affil:Mdy}{
     VSOLJ,
     Kaminishiyamamachi 12-2, Nagasaki, Nagasaki 850-0006, Japan}

\authorcount{affil:Myy}{
     VSOLJ,
     290-383, Ogata-cho, Sasebo, Nagasaki 858-0926, Japan}

\end{center}

\begin{abstract}
\xxinput{abst.inc}
\end{abstract}

   CR Boo is one of the brightest and most famous AM CVn
stars containing a white dwarf and a mass-transferring
helium white dwarf
[for a review of AM CVn stars, see e.g., \citet{sol10amcvnreview}].
CR Boo has received much attention since this object
exhibits dwarf nova-type outbursts
[see e.g., \citet{osa96review}] and its behavior can provide
observational tests for the disk instability theory
in helium disks \citep{tsu97amcvn,sol10amcvnreview,kot12amcvnoutburst}.
CR Boo has been studied by many authors,
starting from the identification of large-amplitude
variations with a quasi-period of four to five days by
\citet{woo87crboo}.  Subsequent research included
\citet{wen94crboo,pat97crboo,pro97crboo,kat00crboo,
kat01crboo,ram12amcvn,hon13crboo,Pdot4,Pdot5,lev15amcvn,
iso16crboo,duf21amcvn,bon22crboo}.

   We used the All-Sky Automated Survey for Supernovae
(ASAS-SN) Sky Patrol data \citep{ASASSN,koc17ASASSNLC} and
the Asteroid Terrestrial-impact Last Alert System
(ATLAS: \cite{ATLAS}) forced photometry \citep{shi21ALTASforced}
and our snapshot CCD photometry
in addition to observations reported to VSOLJ and
VSNET \citep{VSNET}.
Y. M. used a 25-cm RC telescope
with a Canon EOS Kiss X3 DSLR camera.
M. M. used a 25-cm SCT telescope with an SBIG ST-10XME camera.

\begin{figure*}
\begin{center}
\includegraphics[width=14cm]{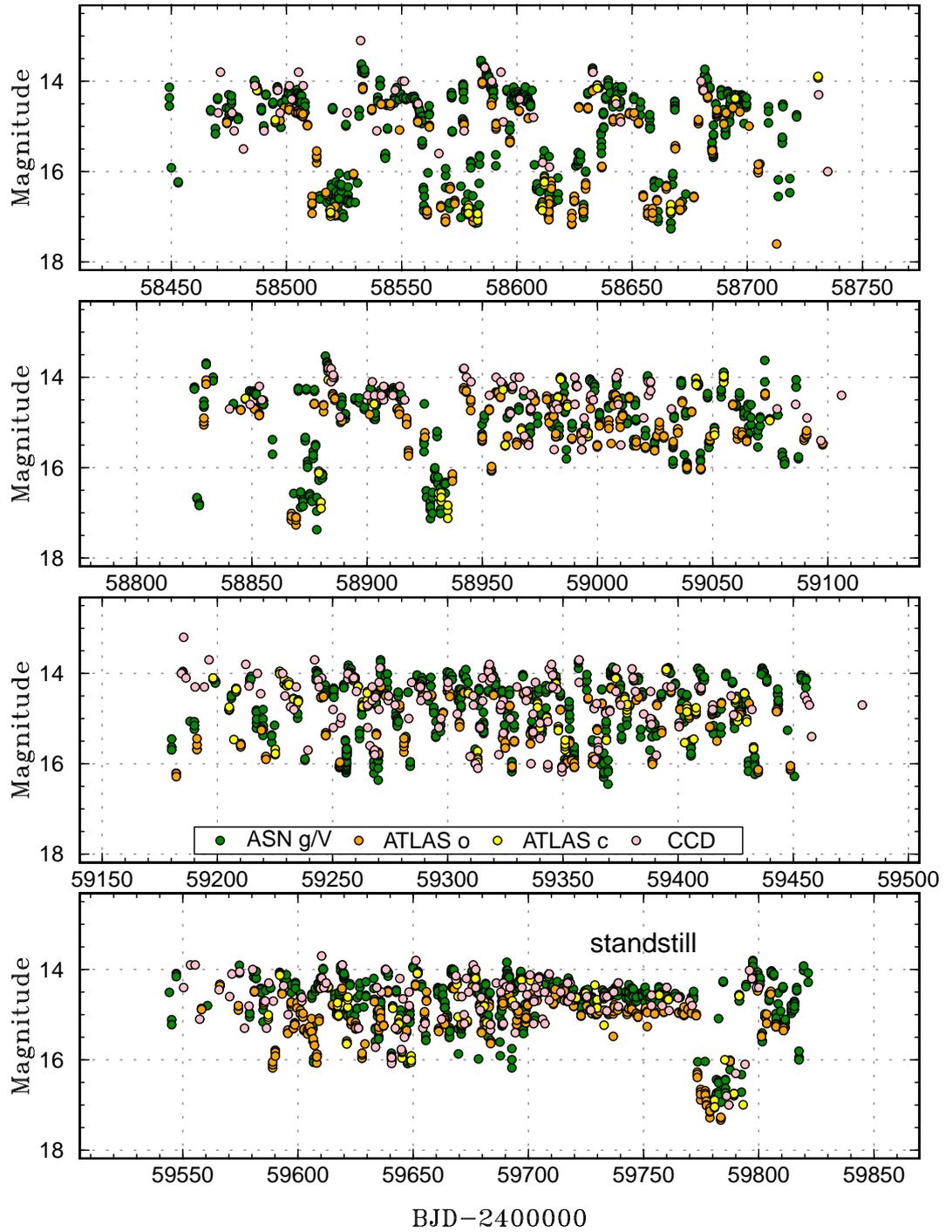}
\caption{
   Light curve of CR Boo in 2018--2022.
   ASAS-SN (ASN) $g$ and ($V$, only one night in the initial part)
   were combined together with $V$ observations reported to
   VSOLJ and VSNET (five observations).
   The category CCD includes unfileted snapshot CCD observations
   mainly by Y.M. and M.M.
   In the first panel (2018--2019 season), CR Boo showed a regular
   pattern of supercycles resembling an ER UMa star among
   hydrogen-rich systems \citep{kat00crboo}.  In the second panel
   (2019--2020), the object initially showed a regular pattern
   as seen in the preceding season, followed by a phase with
   low-amplitude ($\sim$1.5~mag) frequent outbursts with
   increasing cycle length toward the end.
   In the third panel (2020--2021), the object showed a relatively
   regular pattern of outbursts with larger amplitudes
   (up to 2.0~mag) with cycle lengths of 15--20~d.
   This pattern corresponds to the second supercycle described
   in \citet{kat01crboo}.
   In the fourth panel (2021--2022), the object initially showed
   a pattern similar to the 2020--2021 season.
   The amplitudes decreased and entered a standstill at around
   BJD 2459710.  The standstill lasted for $\sim$60~d and
   faded to 17~mag showing short outbursts.
   This faint state lasted for $\sim$20~d.
}
\label{fig:lc}
\end{center}
\end{figure*}

   In figure \ref{fig:lc}, we show the light curve of CR Boo
in the last four seasons.
In the first panel (2018--2019 season), CR Boo showed a regular
pattern of supercycles resembling an ER UMa star among
hydrogen-rich systems \citep{kat00crboo}.
This is probably a typical state in this object.

   In 2019--2020 (second panel of figure \ref{fig:lc}),
the object initially showed
a regular pattern as seen in the preceding season, followed by
a phase with low-amplitude ($\sim$1.5~mag) frequent outbursts with
increasing cycle length toward the end.
Such a pattern is more frequently seen in V803 Cen (sometimes
considered as the CR Boo ``cousin''): see e.g., \citet{duf21amcvn}
figure 1, although \citet{duf21amcvn} apparently did not pay
special attention to this phenomenon and was not seen in CR Boo
during the interval discussed by \citet{duf21amcvn}.
Although this state may be similar to oscillating state as observed
by \citet{pat97crboo} and ``standstill'' with low-amplitude
($\sim$0.5--1 mag) variations reported in V803 Cen
by \citet{kat01v803cen}, the amplitudes were apparently larger
and the long duration of this state suggests a form of
quasi-stable state in helium dwarf novae, which may be
a variety of quasi-periodic variations reported
in \citet{kat01crboo}.

   In the third panel (2020--2021) of figure \ref{fig:lc},
the object showed a relatively
regular pattern of outbursts with larger amplitudes
(up to 2.0~mag) with cycle lengths of 15--20~d.
This pattern corresponds to the second supercycle described
in \citet{kat01crboo}.
In the fourth panel (2021--2022), the object initially showed
a pattern similar to the 2020--2021 season.
The amplitudes decreased and entered a standstill at around
BJD 2459710.  The standstill lasted for $\sim$60~d and
faded to 17~mag showing short outbursts.
This faint state lasted for $\sim$20~d.

\begin{figure*}
\begin{center}
\includegraphics[width=16cm]{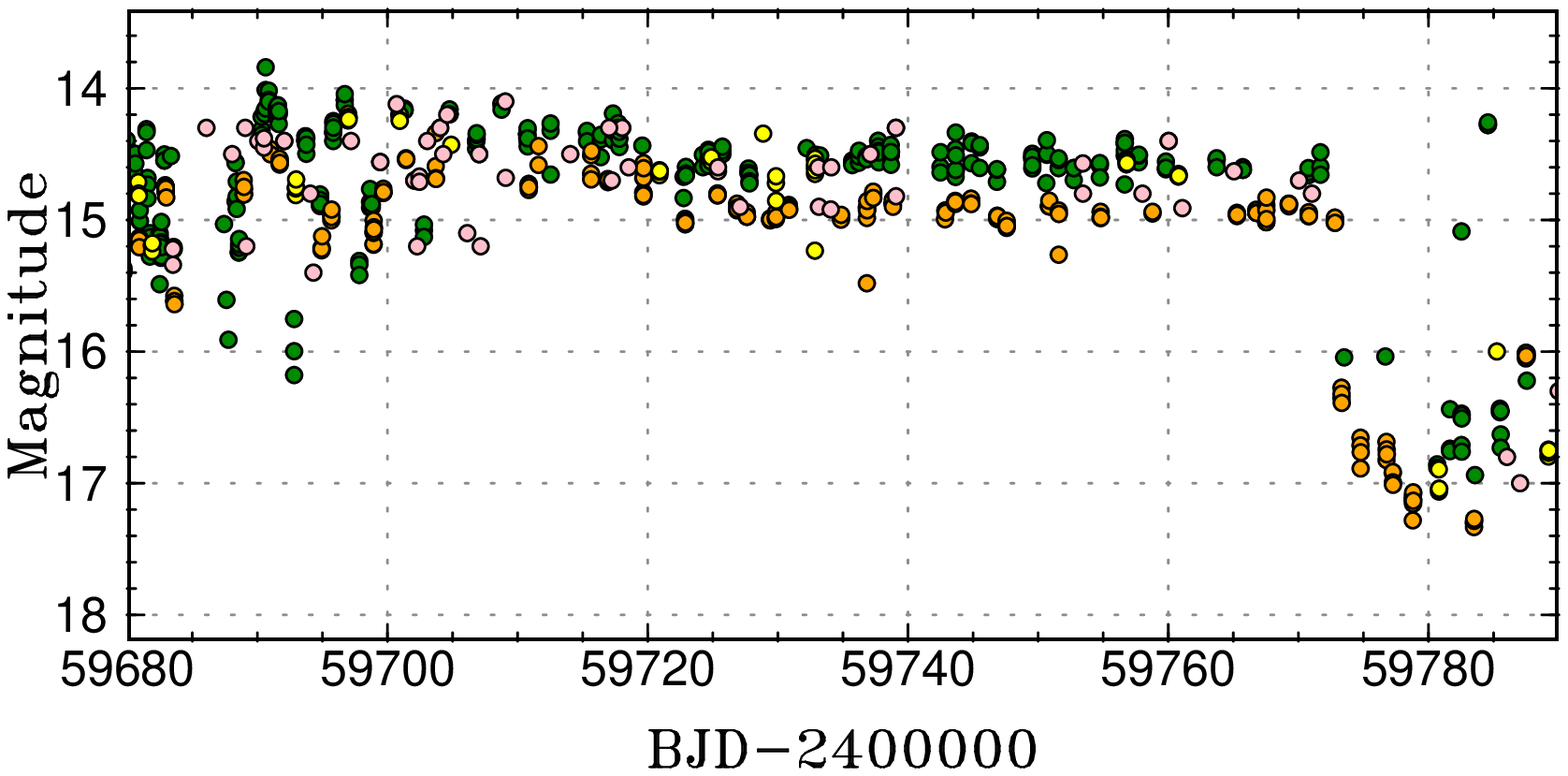}
\caption{
   Enlargement of the light curve including the 2022 standstill.
   The symbols are the same as in figure \ref{fig:lc}.
   The object was essentially almost constant within 0.2~mag
   during the standstill starting from around BJD 2459710,
   as best shown by the ASAS-SN $g$ and ATLAS o data.  
}
\label{fig:lcso}
\end{center}
\end{figure*}

   The details of the standstill in 2022 is shown
in figure \ref{fig:lcso}.
The object was essentially almost constant within 0.2~mag
during the standstill, as best shown by the ASAS-SN $g$ and
ATLAS o data.  Based on the constancy of the brightness for
a long time ($\sim$60~d), we identified this phenomenon
to be a true standstill in this system.
Such a phenomenon was never reported in this system
\citep{pat97crboo,hon13crboo,lev15amcvn,duf21amcvn},
and probably none in other helium dwarf novae
\citep{lev15amcvn,duf21amcvn}.
In figure 1 of \citet{lev15amcvn}, V803 Cen possibly showed
a similar state around JD 2454500--2456700 but only with
a small number of measurements.  The data by ASAS-3 \citep{ASAS3}
and VSNET observations (mainly by Rod Stubbings) of
the corresponding interval could not confirm the constancy
and we consider that this was unlikely a true standstill.

   There was some hint of a standstill-like phenomenon
in CR Boo following a superoutburst and subsequent dips
(figure 3 in \cite{duf21amcvn}).  The same feature was
observed in some superoutbursts of V803 Cen (initial part
of V803 Cen shown in figure 1 of \cite{duf21amcvn}).
This feature was first discovered during observing
campaigns run by VSNET (PI: K. Isogai)
in 2016 and 2017 (the main observers were
Berto Monard, Josch Hambsch, Peter Starr and Gordon Myers),
see e.g., vsnet-alert 19830\footnote{
  $<$http://ooruri.kusastro.kyoto-u.ac.jp/mailarchive/vsnet-alert/19830$>$.
} and 19855\footnote{
  $<$http://ooruri.kusastro.kyoto-u.ac.jp/mailarchive/vsnet-alert/19855$>$.
}.  This feature is similar to damping oscillations seen
in some hydrogen-rich WZ Sge-type dwarf novae following
a superoutburst: ASASSN-15po \citep{nam17asassn15po}
and PQ And \citep{tam21DNespec}.  These phenomena in
V803 Cen and hydrogen-rich WZ Sge-type dwarf novae have much
shorter (less than $\sim$10~d) durations and were preceded by
a superoutburst, and they are usually considered as a part of
the rebrightening phenomenon of WZ Sge stars (see \cite{kat15wzsge}).
The 2022 standstill of CR Boo had a much longer duration and
was not preceded by a superoutburst and is different from
a part of the rebrightening phenomenon.
We therefore consider that CR Boo is a helium Z Cam star
[for classification of cataclysmic variables, see \citet{war95book}].
As in many hydrogen-rich Z Cam stars, the standstill ended
with fading.  The brightness after the standstill was similar
to those after superoutbursts [see also \citet{iso16crboo} for
a study of a typical superoutburst].
The standstill in CR Boo may have had an effect similar to
a superoutburst on the disk -- sweeping out a significant fraction
of the mass stored in the disk as in (hydrogen-rich) SU UMa-type
superoutbursts \citep{osa89suuma}.
Although the presence of superhumps was unfortunately not confirmed
during this standstill, they were likely present considering
the low mass ratio of this system, and the effective removal
of the angular momentum by tidal instability may have worked
just as in an SU UMa-type superoutburst.

   The existence of a standstill in itself may not be
surprising for an object near the thermal stability
\citep{tsu97amcvn,kot12amcvnoutburst}.  In hydrogen-rich
Z Cam stars, a subtle change in the mass transfer is considered
to responsible for standstills \citep{mey83zcam}.
Whether this is also the case in helium dwarf novae
requires further investigation, and, if it is the case,
it would be an interesting question how a degenerate helium
white dwarf (without a solar-type magnetic cycle as expected
for secondaries in hydrogen-rich Z Cam stars) can produce
such a variation in mass transfer.  If it is not the case,
it would be an interesting question how the disk can
produce various states without variation in the mass transfer
from the secondary.

\section*{Acknowledgements}

This work was supported by JSPS KAKENHI Grant Number 21K03616.

We are grateful to the ATLAS, ASAS-SN and ASAS-3 teams
for making their data available to the public.
We are also grateful to observers (particularly Masao Funada
and Gary Poyner) who provided data of CR Boo in 2018--2022,
and observers who contributed to campaigns of V803 Cen
(particularly Berto Monard, Josch Hambsch, Peter Starr,
Gordon Myers and Rod Stubbings).

This work has made use of data from the Asteroid Terrestrial-impact
Last Alert System (ATLAS) project. The Asteroid Terrestrial-impact
Last Alert System (ATLAS) project is primarily funded to search for
near earth asteroids through NASA grants NN12AR55G, 80NSSC18K0284,
and 80NSSC18K1575; byproducts of the NEO search include images and
catalogs from the survey area. This work was partially funded by
Kepler/K2 grant J1944/80NSSC19K0112 and HST GO-15889, and STFC
grants ST/T000198/1 and ST/S006109/1. The ATLAS science products
have been made possible through the contributions of the University
of Hawaii Institute for Astronomy, the Queen's University Belfast, 
the Space Telescope Science Institute, the South African Astronomical
Observatory, and The Millennium Institute of Astrophysics (MAS), Chile.

\section*{List of objects in this paper}
\xxinput{objlist.inc}

\section*{References}

We provide two forms of the references section (for ADS
and as published) so that the references can be easily
incorporated into ADS.

\renewcommand\refname{\textbf{References (for ADS)}}

\newcommand{\noop}[1]{}\newcommand{\hyphalt}{-}

\xxinput{crboossaph.bbl}

\renewcommand\refname{\textbf{References (as published)}}

\xxinput{crbooss.bbl.vsolj}

\end{document}